# Simulation Study of Hemispheric Phase-Asymmetry in the Solar Cycle

D. Shukuya

Institute for Space-Earth Environmental Research, Nagoya University

Furo-cho Chikusa-ku, Nagoya, Aichi 4648601, Japan.

K. Kusano

Institute for Space-Earth Environmental Research, Nagoya University

Furo-cho, Chikusa-ku, Nagoya, Aichi 4648601, Japan.

kusano@nagoya-u.jp

Abstract

Observations of the sun suggest that solar activities systematically create north-south hemispheric asymmetries. For instance, the hemisphere in which the sunspot activity is more active tends to switch after the early half of each solar cycle. Svalgaard & Kamide (2013) recently pointed out that the time gaps of polar field reversal between the north and south hemispheres are simply consequences of the asymmetry of sunspot activity. However, the mechanism underlying the asymmetric feature in solar cycle activities is not yet well understood. In this paper, in order to explain the cause of the asymmetry from the theoretical point of view, we investigate the relationship between the dipole- and quadrupole-type components of the magnetic field in the solar cycle using the mean-field theory based on the flux transport dynamo model. As a result, we found that there are two different attractors of the solar cycle, in which either the north or the south polar field is first reversed, and that the flux transport dynamo model well explains the phase-asymmetry of sunspot activity and the polar field reversal without any ad hoc source of asymmetry.

1. INTRODUCTION

Various kinds of hemispheric asymmetry in the northern and southern hemispheres have been observed in solar magnetic activities, e.g., sunspot numbers, sunspot area, flares, prominences, faculae, coronal brightness, and the time of polar magnetic field reversal (see e.g., Maunder 1890, 1904, Waldmeier 1971, Roy 1977, and Babcock 1959). According to Waldmeier (1971), for instance, the phase of sunspot cycle 19 was shifted between the northern and the southern hemispheres, such that the sunspot cycle in the southern hemisphere reached the maximum approximately one year earlier than that in the northern hemisphere. In addition, it was reported that the polar magnetic field reversal at one pole is delayed for one or two years compared with the other pole in every solar cycle (Babcock 1959, Li 2009, Murakozy & Ludmany 2012, Shiota et al. 2012, and Svalgaard & Kamide 2013). Based on the observation of sunspot groups during the last 12 solar cycles, Li (2009) and Murakozy & Ludmany (2012) pointed out that the leading hemisphere, in which the magnetic activities are more active than the other in the early half-phase of each cycle, may switch after approximately an eight-sunspot-cycle period. These observational results imply that certain systematic processes in addition to a stochastic process govern the asymmetries of the solar cycle. However, the mechanisms underlying this have not yet been clearly explained.

The solar magnetic field is considered to be sustained by the dynamo action in the interior of the sun (Parker 1955). After the helioseismology revealed the averaged distribution of the large-scale internal velocity field, i.e., the meridional circulation and the differential rotation (Thompson et al. 2003), the flux transport dynamo model was proposed (Choudhuri et al. 1995, Dikpati & Charbonneau 1999, Nandy & Choudhuri 2002, and Chatterjee, Nandy, & Choudhuri (hereafter CNC) 2004) as a promising model to explain certain features of solar magnetic activities, such as the 11-year cycle, the butterfly diagram, the poleward migration of surface field, and polarity reversals of the polar field. Since the computation of the flux transport model is not numerically demanding, numerical simulations based on this model are widely used as a tool to study the variation in the solar cycle.

In previous studies, several numerical models ware developed to explain the asymmetric features of the solar cycle (e.g., DeRosa et al. 2012, Brun et al. 2013, Belucz et al. 2013, Belucz & Dikpati 2013, Olemskoy & Kitchatinov 2013, and Shetye et al. 2015). DeRosa et al. (2012) and Brun et al. (2013) showed that a certain degree of asymmetric modulation to the source of the poloidal field or the meridional circulation causes the interaction between the primary family and the secondary family of the magnetic field and produces the time gaps of the polar magnetic field reversals between the north and the south poles. The primary family is a dipole-type part, which is the spherical harmonic functions of the surface magnetic field $B_l^0$ with $l = 1, 3, 5, 7, \cdots$, and the secondary family is a quadrupole-type part, which is $B_l^0$ with $l = 2, 4, 6, 8, \cdots$. Belucz et al. (2013) showed that the dynamo operates mostly independently in the northern and the southern hemispheres by studying the features of the dynamo operating with the source terms of poloidal field of different amplitudes between the different hemispheres. Belucz & Dikpati (2013) showed that when the amplitude or pattern of the meridional circulation changed only for the southern hemisphere, the dynamo period, the shape of the butterfly diagram, the strength of the polar and toroidal fields, and the phase relations between the polar and toroidal fields changed almost exclusively for the southern hemisphere. Olemskoy & Kitchatinov (2013) introduced the fluctuating poloidal field source term into the mean-field dynamo model using a smoothly varying random function of time and latitude. As a consequence, it was shown that the fluctuations violate hemispheric symmetry of the dynamo field. When the deviations from dipole parity are large, the model shows weak magnetic cycles, in which the large asymmetry of magnetic activity appears like the sunspot in grand minima. Passos et al. (2013) also studied the origin of hemispheric asymmetry and cycle amplitude modulation by introducing independent stochastic fluctuations in the two hemispheres using the mean-field dynamo model, and they demonstrated many types of hemispheric asymmetries, including grand minima and failed grand minima where only one hemisphere enters a quiescent state. Shetye et al. (2015) discussed the contribution of active-region fluxes and their hemispheric asymmetries. They introduced the asymmetric active-region inflow effect to the meridional circulation and found that the inflow can affect peak amplitude of the solar cycle. Observations by local helioseismology indicate that the meridional motions of active regions generate inflows, which are meridional converging flows into active regions at the solar surface from lower and higher latitudes (Gizon & Birch 2005).

These previous studies indicate that the symmetric properties of the dynamo are obviously affected by the asymmetric flow or magnetic field, which is ad hoc imposed in the dynamo equation. However, the cause of this asymmetry is still open to question. The objective of this paper is to elucidate whether the asymmetric properties in the solar dynamo are capable of being spontaneously created even though the flow is symmetric with respect to the equator. When the flow is symmetric the dynamo equation is known to have two different types of solutions, i.e., the so-called dipole-type and quadrupole-type families (Jennings & Weiss 1991 and Nishikawa & Kusano 2008). Since these fields have different parities of hemispheric symmetry, their mixing might cause the asymmetries. Our study was originally motivated by this property of the dynamo equation. Since dynamo is a nonlinear process in which flow and magnetic field mutually interact, it is not easy to elucidate what is the fundamental cause of the asymmetric property. However, if some asymmetry is created even though any external conditions (e.g., flow, source terms, and diffusion) are symmetric, we could conclude that the asymmetric property of the dynamo is inherent in the dynamo equation. To achieve this, we developed a standard flux transport dynamo code based on the SURYA code (CNC 2004) and carefully analyzed the asymmetric features of the solar magnetic activities. Finally, we will propose that the solar cycle may spontaneously create the hemispheric asymmetry.

## 2. SIMULATION MODEL

### 2.1. Basic Equations

We solve the axisymmetric mean-field kinematic dynamo equations on a meridional plane of a rotating spherical shell ($R_b = 0.55 R_\odot \leq r \leq R_\odot$, $0 \leq \theta \leq \pi$, where $R_b$, $R_\odot$, $r$, and $\theta$ denote the radius of the bottom boundary, the solar radius, the radial coordinate, and the colatitude, respectively). The shell is rotating with an angular velocity $\boldsymbol{\Omega} = \Omega \boldsymbol{e}_z$, where $\boldsymbol{e}_z$ is a unit vector along the rotation axis. Hereafter, $\boldsymbol{e}_i$ denotes a unit vector along the $i$ coordinate.

Under the axisymmetric approximation, the azimuthally averaged magnetic field $\boldsymbol{B}(r,\theta,t)$ and velocity field $\boldsymbol{V}(r,\theta)$ can be decomposed into the poloidal and toroidal components, respectively:

$$\begin{aligned}\boldsymbol{B}(r,\theta,t) &= \boldsymbol{B}_p(r,\theta,t) + B(r,\theta,t)\boldsymbol{e}_\phi \\ &= \nabla\times[A(r,\theta,t)\boldsymbol{e}_\phi] + B(r,\theta,t)\boldsymbol{e}_\phi,\end{aligned} \quad (1)$$

$$\begin{aligned}\boldsymbol{V}(r,\theta) &= \boldsymbol{v}_p(r,\theta) + r\sin\theta\ \Omega(r,\theta)\boldsymbol{e}_\phi \\ &= v_r \boldsymbol{e}_r + v_\theta \boldsymbol{e}_\theta + r\sin\theta\ \Omega(r,\theta)\boldsymbol{e}_\phi,\end{aligned} \quad (2)$$

in which the poloidal magnetic field $\boldsymbol{B}_p(r,\theta,t)$ is described by the toroidal ($\phi$) component of the vector potential $A(r,\theta,t)$ as $\boldsymbol{B}_p(r,\theta,t) = \nabla\times[A(r,\theta,t)\boldsymbol{e}_\phi]$. The vectors, $B(r,\theta,t)\boldsymbol{e}_\phi$ and $\boldsymbol{v}_p(r,\theta)$, are the toroidal component of magnetic field and the poloidal component of velocity field (meridional flow), respectively. Using Equations (1) and (2), the basic equations for the so-called $\alpha\omega$ dynamo are

$$\frac{\partial B}{\partial t} + \frac{1}{r}\left[\frac{\partial}{\partial r}(r v_r B) + \frac{\partial}{\partial \theta}(v_\theta B)\right] \\ = \eta_t\left(\nabla^2 - \frac{1}{s^2}\right)B + \frac{1}{r}\frac{d\eta_t}{dr}\frac{\partial}{\partial r}(rB) + s(\boldsymbol{B}_P\cdot\boldsymbol{\nabla})\Omega, \quad (3)$$

$$\frac{\partial A}{\partial t} + \frac{1}{s}(\boldsymbol{v}_P \cdot \boldsymbol{\nabla})(sA) = \eta_p \left(\nabla^2 - \frac{1}{s^2}\right)A + \alpha B, \tag{4}$$

$$\eta_p(r) = \eta_{RZ} + \frac{\eta_{SCZ}}{2}\left[1 + \mathrm{erf}\left(\frac{r - r_{BCZ}}{d_t}\right)\right], \tag{5}$$

$$\eta_t(r) = \eta_{RZ} + \frac{\eta_{SCZ1}}{2}\left[1 + \mathrm{erf}\left(\frac{r - r'_{BCZ}}{d_t}\right)\right] + \frac{\eta_{SCZ}}{2}\left[1 + \mathrm{erf}\left(\frac{r - r_{TCZ}}{d_t}\right)\right]. \tag{6}$$

These are derived from the magnetic induction equation in magnetohydrodynamics (MHD), where $s = r\sin\theta$, and $\eta_p$ and $\eta_t$ are the coefficients of the net magnetic diffusivity for the poloidal and toroidal components of the magnetic field, respectively. In this paper, we prescribe the angular velocity $\Omega$, meridional flow $\boldsymbol{v}_p$, diffusivity coefficient $\eta_t$, and coefficient $\alpha$ in the same way that was used in CNC (2004) and Choudhuri et al. (2005). The maximum amplitude of meridional flow speed is 29.0 [m/s]. The term of $\alpha B$ on the right-hand side of Equation (4) plays the role of the source for the poloidal magnetic field near the solar surface due to the tilt of active regions called the Joy's law. This is called the Babcock–Leighton $\alpha$-effect, in which the tilted bipole reproduces the poloidal field near the solar surface (Babcock 1961 and Leighton 1969).

   The previous studies (CNC 2004 and Hotta & Yokoyama 2010) showed that the dipole-type solution switches to the quadrupole-type solution, for example, when the diffusivity coefficient $\eta_p$ increases and/or the depth of the strong diffusivity layer near the surface becomes thinner. In this paper, we focus on the diffusivity coefficient ($\eta_{SCZ}$) and survey the dependency on $\eta_{SCZ}$ for $10^{11} - 10^{13}$ cm$^2$s$^{-1}$.

   We normalize time so that the rotation period at the equator on the surface is 25 (corresponding to the number of days) (Thompson et al. 2003) and magnetic field by the critical field $B_c$, which will be explained later in detail.

2.2. Decomposition of Hemispheric Symmetry

   In order to analyze the asymmetric features of the dynamo action, we introduce the following procedure to decompose the equations into two groups of different symmetries. In the spherical coordinate, an arbitrary scalar function $y(\theta)$ of the colatitude $\theta$ can be uniquely decomposed into the two functions of different parities,

$$y(\theta) = y^s(\theta) + y^a(\theta) \tag{7}$$

where $y^s(\theta)$ is symmetric with respect to the equator and $y^a(\theta)$ is anti-symmetric:

$$\begin{aligned} y^s(\theta) &\equiv [y(\theta) + y(\pi - \theta)]/2, \\ y^a(\theta) &\equiv [y(\theta) - y(\pi - \theta)]/2. \end{aligned} \tag{8}$$

In the same way, an arbitrary vector $\boldsymbol{Y}(\theta)$ can also be uniquely decomposed into the two vectors,

$$\boldsymbol{Y} = \boldsymbol{Y}^s + \boldsymbol{Y}^a, \tag{9}$$

where

$$\begin{aligned} \boldsymbol{Y}^s &\equiv y_r^s \boldsymbol{e}_r + y_\theta^a \boldsymbol{e}_\theta + y_\phi^s \boldsymbol{e}_\phi, \\ \boldsymbol{Y}^a &\equiv y_r^a \boldsymbol{e}_r + y_\theta^s \boldsymbol{e}_\theta + y_\phi^a \boldsymbol{e}_\phi. \end{aligned} \tag{10}$$

Moreover, the conversion of symmetric properties by vector operations is prescribed (for full details, see Section III of Nishikawa & Kusano 2008). Using the vector formula, we decompose Equation (3) and (4) into the two different groups of dynamo equations,

$$\begin{aligned} \frac{\partial}{\partial t} B^s &+ \frac{1}{r}\left[\frac{\partial}{\partial r}(rv_r^s B^s) + \frac{\partial}{\partial \theta}(v_\theta^a B^s)\right] \\ &= \eta_t \left(\nabla^2 - \frac{1}{s^2}\right) B^s + \frac{1}{r}\frac{d\eta_t}{dr}\frac{\partial}{\partial r}(rB^s) + s(\boldsymbol{B}_P^s \cdot \boldsymbol{\nabla})\Omega^s, \end{aligned} \tag{11}$$

$$\frac{\partial}{\partial t} A^a + \frac{1}{s}(\boldsymbol{v}_P^s \cdot \boldsymbol{\nabla})(sA^a) = \eta_p \left(\nabla^2 - \frac{1}{s^2}\right) A^a + \alpha^a B^s, \tag{12}$$

for $B^s$ and $A^a$ and

$$\frac{\partial}{\partial t}B^a + \frac{1}{r}\left[\frac{\partial}{\partial r}(rv_r^s B^a) + \frac{\partial}{\partial \theta}(v_\theta^a B^a)\right]$$
$$= \eta_t\left(\nabla^2 - \frac{1}{s^2}\right)B^a + \frac{1}{r}\frac{d\eta_t}{dr}\frac{\partial}{\partial r}(rB^a) + s(\boldsymbol{B}_P^a \cdot \boldsymbol{\nabla})\Omega^s, \quad (13)$$

$$\frac{\partial}{\partial t}A^s + \frac{1}{s}(\boldsymbol{v}_P^s \cdot \boldsymbol{\nabla})(sA^s) = \eta_p\left(\nabla^2 - \frac{1}{s^2}\right)A^s + \alpha^a B^a, \quad (14)$$

for $B^a$ and $A^s$, if velocities $\boldsymbol{v}_p$ and the rotation $\Omega$ are symmetric. These equations clearly indicate that the symmetric component of the vector equations is constructed only with $B^s$ and $A^a$ and the anti-symmetric component is constructed only with $B^a$ and $A^s$. Therefore, these components are mutually independent as long as the velocity is symmetric. The symmetric and anti-symmetric components of the magnetic field correspond to the so-called quadrupole-type and dipole-type families, respectively (Jennings & Weiss 1991 and Nishikawa & Kusano 2008). These would couple with each other if and only if the anti-symmetric velocity field exists.

We numerically solved Equations (11) to (14) for $0 < \theta < \pi/2$, although they are identical to Equation (3) and (4) for $0 < \theta < \pi$, in order to avoid the incorrect mixing between the two families in the solution of Equation (3) and (4) due to numerical noise. In our calculations, although the coordinate $\theta$ is defined only between 0 and $\pi/2$, the magnetic field in the northern and southern hemispheres is easily reproduced by $B^s + B^a$ and $B^s - B^a$, respectively, and thus the simulation box covers the entire sphere.

The numerical algorithm was the same as in the previous study (CNC 2004), and we used the alternating direction implicit (ADI) method. We handled the diffusion terms through a centered-difference scheme and the advection terms through the Lax–Wendroff scheme. The grid numbers ware 141 for both $0.55R_\odot \leq r \leq R_\odot$ and $0 \leq \theta \leq \pi/2$. The numerical convergence was checked with runs using twofold-greater grid numbers.

2.3. Boundary Conditions

The boundary conditions are as follows. On the rotation axis ($\theta = 0$), we have

$$A = 0, \quad B = 0. \quad (15)$$

The bottom boundary $(r = R_b)$ condition is

$$A = 0, \quad B = 0. \quad (16)$$

At the top boundary $(r = R_\odot)$, we impose the condition

$$\left(\nabla^2 - \frac{1}{s^2}\right) A = 0, \quad B = 0, \tag{17}$$

in which $A$ matches smoothly to an exterior potential field solution (Dikpati & Choudhuri 1994). At the equator $(\theta = \pi/2)$, the symmetric component satisfies

$$A^a = 0, \quad \frac{\partial}{\partial \theta} B^s = 0, \tag{18}$$

and the anti-symmetric component satisfies

$$\frac{\partial}{\partial \theta} A^s = 0, \quad B^a = 0. \tag{19}$$

### 2.4. The Magnetic Buoyancy Effect

The magnetic buoyancy may cause a strong toroidal magnetic field to erupt to the upper region. Helioseismology has revealed that there is a region named tachocline at the base of the convection zone, which has large shear of the angular rotation. It is thought that the tachocline generates the toroidal magnetic field more efficiently than other regions and the magnetic buoyancy occurs mainly there. Therefore, we implemented the magnetic buoyancy effect by the following procedure. First, we convert $B^s$ and $B^a$ to the toroidal field $B$, and search for the place where $B$ exceeds a critical value, $B_c$, above the bottom of the convection zone $(r = 0.71 R_\odot)$ every $\tau = 10 \ [days]$. Second, if the toroidal field is larger than $B_c$, we move half of the magnetic flux at this position to just below the solar surface and the same latitude based on the method developed by Nandy & Choudhuri (2001). Finally, we redefine $B^s$ and $B^a$.

### 3. RESULTS

#### 3.1. Phase Relations between the Symmetric and Anti-symmetric Components

We show the result of our simulation in Figure 1, which describes the time evolution of the radial magnetic field for the symmetric and anti-symmetric components at the north pole. These cyclic variations correspond to the Hale cycle, which is the solar magnetic cycle with an average duration of 22 years. The initial condition is given by the combination of symmetric and anti-symmetric components, which are premade by the preceding calculations. In this case, the amplitudes and the phases for the each component are initially the same. We found that their phases were gradually shifted during the first five cycles. Then, the phase difference is fixed so that the symmetric component precedes the anti-symmetric component by about a quarter period (~90 degrees).

To investigate how the phase difference depends on the initial condition, we performed the calculations for 22 different simulations in which the initial phase difference $\Delta\psi$ was varied within the range $-180° \leq \Delta\psi \leq 180°$, by varying $\psi_s$ with respect to $\psi_a$. The values, $\psi_s$ and $\psi_a$, are the phases for the symmetric and anti-symmetric components and the defined using the following equation:

$$\psi_i = \tan^{-1}\left(\frac{B_r^i}{dB_r^i/dt}\right), \qquad (i = s, a). \tag{20}$$

Figure 2 shows the time evolution of $\Delta\psi$ for the various simulations. The results indicate that the phase differences for the initial $\Delta\psi$ below 0° to −180° fall into $\Delta\psi = -90°$ and those for the initial $\Delta\psi$ over 0° to 180° fall into $\Delta\psi = 90°$. This implies that the phase relations between the two components have only two attractors ($\Delta\psi = 90°$ and $\Delta\psi = -90°$, respectively) and that either of them is always achieved depending on the initial conditions of $\Delta\psi$. The two attractors may occupy a half-area of parameter space for $\Delta\psi$.

3.2. Cyclic Behaviors of Magnetic Field in Each Attractor

Figures 3 and 4 show the time evolutions of the magnetic field for several cycles after falling into each attractor with $\Delta\psi = 90°$ and $\Delta\psi = -90°$, respectively, to inspect behaviors within a cycle. The dashed line in the top panels show the total counts of grids where the toroidal magnetic field exceeds the critical intensity for the magnetic buoyancy effect. The counts represent the sunspot activity. The solid lines indicate the count difference between the northern and southern hemispheres. The value is positive when the activity is higher in the northern hemisphere and negative when the activity is higher in the southern hemisphere, as shown in the second panel which indicates the time evolution of the number of grids exceeding the critical buoyancy in the northern (solid line) and southern (dashed line) hemispheres, respectively. The third panel shows the time evolution of the radial magnetic field $B_r$ at the poles: the northern field $B_r^{Nor}$ is indicated by the solid line and the southern field $B_r^{Sou}$ by the dashed line. The fourth panel shows the time evolution of the radial magnetic field $B_r$ of the symmetric (solid line) and anti-symmetric (dashed line) components at the north pole, as in Figure 1. The contour maps at the bottom of each figure show the contour of the vector potential $A$ at the times corresponding to the vertical dashed lines in the top three panels. The solid and dashed contours represent the magnetic field lines directed by arrows.

The cycle activity observed in Figure 3 is summarized as follows:

1) Cycle from $t = 17$ to $30$ $[years]$: First, if we focus on a cycle from $t = 17$ to $30$ $[years]$, we see that the sunspot activity in the northern hemisphere is higher than that in the southern hemisphere in the first half of this period ($t = 17$ to $23$ $[years]$), while the southern hemisphere is more active in the second half. This feature is repeated in every cycle in this case. Second, focusing on the second panel and the bottom contour maps, the poloidal magnetic field at $t = 17$ $[years]$ has a dipole-type configuration, in which the north pole has a positive field and the south pole has a negative field. Then, the magnetic field at the north pole reverses at around $t = 22$ $[years]$, and the poloidal magnetic field changes to the quadrupole-type configuration, as seen in the contour map at $t = 23$ $[years]$. This quadrupole-type configuration persists until the magnetic field at the south pole reverses at around $t = 26$ $[years]$.

2) Cycle from $t = 30$ to $42$ $[years]$: Then, the dipole-type configuration, in which the polarity is opposite to that at $t = 17$ $[years]$ reappears. After the magnetic field at the north pole reverses at around $t = 33$ $[years]$, it switches to the quadrupole-type configuration again, and after the magnetic field at the south pole reverses at around $t = 37$ $[years]$, the dipole-type configuration whose polarity is the same as at $t = 17$ $[years]$ recovers.

These results clearly indicate that the magnetic field $B_r$ always reverses at the north pole first, and the polarity reversal at the south pole follows. On the other hand, in the attractor for $\Delta\psi = -90°$, the sunspot activity in the southern hemisphere is higher than that in the northern hemisphere in the first half of the cycle, and the polarity reversal at the south pole always precedes the north pole, as seen in Figure 4.

Base on the results above, we can conclude that the flux transport dynamo solution spontaneously generates the asymmetric behavior of the solar cycle, in which the polar field is reversed for the hemisphere that is more active in the early phase of the cycle compared to the other hemisphere.

3.3. Dependency on the Magnetic Reynolds Number $Rm$

We also analyzed the dependence of dynamo activity on the magnetic Reynolds number $Rm$ defined as

$$Rm = \frac{V_0 R_\odot}{\eta_{SCZ}}. \tag{21}$$

Here, the typical values were used for the solar radius $R_\odot$ and the speed difference of differential rotation between the equator ($\theta = 90°$) and polar area ($\theta = 10°$) $V_0 = 1.759 \times 10^3\ ms^{-1}$. We surveyed the dependency on $Rm$ by changing $\eta_{SCZ}$.

The long-term evolution of the amplitude ratio defined as

$$\chi = \frac{B^s_{r,max} - B^a_{r,max}}{B^s_{r,max} + B^a_{r,max}}, \tag{22}$$

is plotted in Figure 5, where $B^s_{r,max}$ and $B^a_{r,max}$ are the maximum values of the radial magnetic field at the north pole in one sunspot cycle for the symmetric and anti-symmetric components, respectively. We performed the simulations for various $Rm$ values, in which the initial condition consists of the two components of the same amplitude, i.e. $\chi = 0$ initially. Figure 5 shows that the ratio $\chi$ gradually changes to different states. When $Rm > 6.06 \times 10^3$, $\chi$ increases, while $\chi$ decreases for $Rm < 6.06 \times 10^3$. This means that the final solution is switched from the anti-symmetric to the symmetric at $Rm \approx 6.06 \times 10^3$, as plotted in Figure 6, which shows the relationship between $Rm$ and $\chi$ at $t = 15000\ [years]$. The critical resistivities $\eta_p$ and $\eta_t$ corresponding to $Rm = 6.06 \times 10^3$ are 2.02x10$^{12}$ [cm$^2$/s] and 0.04x10$^{12}$ [cm$^2$/s], respectively.

Chatterjee et al. (2004) explored a two-dimensional kinematic solar dynamo model in a full sphere and concluded that the dipolar mode is preferred when certain reasonable conditions are satisfied, Thereafter, Yeates et al. (2008) revealed that the nature of the dynamo changes from advection dominated to diffusion dominated for different relative choices of turbulent diffusivity and flow speed (i.e, Reynolds number). Hotta & Yokoyama (2010) also investigated the dependence of the solar magnetic parity between the hemispheres on the turbulent diffusivity and the meridional flow by means of axisymmetric kinematic dynamo model, and they showed that the stronger diffusivity near the surface is more likely to cause the magnetic field to be a dipole. These results are qualitatively consistent with our result.

Figure 7 shows that the ratio $\chi$ is also related to the lag time $\Delta t_{lag}$ between the polar field reversals at the north and south poles. When $\chi = -1$, the anti-symmetric component governs the solution, and the polar fields are reversed simultaneously at both poles ($\Delta t_{lag} = 0$). However, the lag time $\Delta t_{lag}$ increases with $\chi$ up to a half-period cycle when $\chi = 0$. Finally, when $\chi = 1$, the symmetric component dominates and the lag time $\Delta t_{lag}$ reaches the cycle period. This means that the reversal again occurs simultaneously between the poles.

According to DeRosa et al. (2012), the primary-family of magnetic field (corresponding to the anti-symmetric component) attains amplitudes of about 25% of the secondary-family (corresponding to the symmetric component) according to the observation. This may correspond to about $\chi = -0.6$, and Figure 7 indicates that the lag time $\Delta t_{lag}$ for this is $\Delta t_{lag} \approx 1.7 \ [years]$. This result is almost consistent with the observation of polar field reversal mentioned by Babcock (1959) and Svalgaard & Kamide (2013).

4. DISCUSSION

Newton & Milsom (1955), Waldmeier (1971), Li (2009), and Murakozy & Ludmany (2012) derived various indices for the degree of the phase lag in the solar magnetic cycle between the northern and southern hemispheres, using the observational data of sunspot number, the number of sunspot groups, and the area of sunspots. They used these indices to investigate the variation of the phase lag. As summarized in Table 1, these results showed that the leading hemisphere in the solar cycle tends to be maintained for several cycles, and also that it switched to another hemisphere, e.g., on the cycles 11/12, 15/16, and 19/20. Therefore, they suggested that a 4+4 cycle period might exist for the phase lag of the hemispheric cycles.

The observation that the leading hemisphere tends to be maintained is consistent with our result for the two attractors of the solar cycle, in which either the northern or southern hemispheres leads the cycle. However, our simulation cannot explain the periodic switch of the leading hemisphere because, if the cycle falls into an attracter, the solution will never escape from there.

For the attractors, the cycle periods of the dipole and quadrupole-type components coincide and the lag between their phases is locked. In the kinematic dynamo model, however, the two different components have different eigenvalues corresponding to the cycle period. It indicates that some nonlinear effects, which couple the different components, are needed to make the attractors. In our model, only the magnetic buoyancy effect plays this role. In the solar convection zone, however, it is likely that more complicated nonlinear effects work to modulate the cycle period (Cameron et al. 2014).

Although it is beyond the scope of this paper to determine what causes the transition between the attractors of solar cycles, let us discuss what kind of modulation might be consistent with the observed properties. Here, we assume the following: the fundamental features of the attractor are sustained; the two components, $B^s$ and $B^a$, have a common cycle frequency $\omega$; and the phase lag is fixed at $\pi/2$:

$$B^s = M^s \sin(\omega t + \pi/2), \tag{23}$$
$$B^a = M^a \sin(\omega t). \tag{24}$$

On the other hand, we introduce the competitive modulation of a four-cycle period between the two components,

$$M^s = \mu \sin(\omega t/4), \tag{25}$$
$$M^a = 1 - M^s, \tag{26}$$

where $\mu$ is the amplitude of modulation. In fact, Nishikawa & Kusano (2008) revealed that the dipole-type component weakened when the quadrupole-type component grows in their long-term MHD simulation of a quickly rotating spherical shell dynamo. The competitive relationship in Equation (26) is a simple model for this competitive relationship between the two components.

Some results for the modulated cycle are shown in Figure 8, where the primary cycle $\pi/\omega$ and the modulation amplitude $\mu$ are assumed to be $10\ [years]$ and $0.25$, respectively. In this figure, the upper panel shows the evolution of magnetic activity in the northern and southern hemispheres, and the lower panel indicates the amplitude of the symmetric and anti-symmetric components. It shows that the leading hemisphere switches every four cycles, because the sign of the symmetric component $M^s$ is reversed. Though this mathematical experiment is too simple to simulate the realistic solar cycle, it implies that some possibility exists that the variability of the symmetric component might cause the transition of attractors.

## 5. CONCLUSIONS

In order to investigate the mechanism of the phase-asymmetry of the solar cycle between the different hemispheres, we performed dynamo simulations based on the flux transport dynamo model. Our simulation code was devised to precisely calculate the hemispheric symmetricity based on the SURYA code (CNC 2004). As a result, we have shown that the flux transport dynamo model explains the phase asymmetry well. Thus, the phase asymmetry is an inherent property of the cyclic dynamo solutions, and it appears without any ad hoc source of asymmetry in the magnetic field and velocity (e.g. the differential rotation and the meridional circulation). The dynamo solutions of the solar cycle have two attractors, in which the phases of the quadrupole-type and dipole-type components differ by $\Delta\psi = \pm 90°$. When $\Delta\psi = 90°$ ($\Delta\psi = -90°$), the cycle of the northern (southern) hemisphere leads the cycle of other hemisphere. The lag time of the phases between the different hemispheres, $\Delta t_{lag}$, is determined by the amplitude ratio of dipole and quadrupole components $\chi$. The ratio $\chi$ of the attractors depends on the magnetic Reynolds number $R_m$ for the solar convection zone. Within the range of $3.71 \times 10^3 \leq R_m \leq$

$1.02\times10^4$, the dynamo switches from the dipole-dominate solution to the quadrupole-dominate solution.

      Our results are consistent with the observation that the leading hemisphere in the solar cycle tends to be persistent for several cycles. However, our model cannot explain the observation that the leading hemisphere switches at about every four cycles nor the amplitude-asymmetry of each solar cycle between the northern and southern hemispheres. Although the switch of the leading hemisphere may be understood as the transition from one attractor to the other, it is still elusive what kind of nonlinear effects may cause this transition. In order to elucidate this issue, we need to develop a more sophisticated model that can handle the feedback effect of the dynamo field.

ACKNOWLEDGEMENTS

Our simulation code was devised based on the solar dynamo code SURYA developed by professor A. R. Choudhuri and his co-researchers at the Indian Institute of Science, Bangalore. This work was supported by a Grant-in-Aid for the Japan Society for the Promotion of Science (JSPS) Fellows (Grant No.: 26-4779) and MEXT/JSPS KAKENHI 15H05816, 25287051, and 25247014. D. Shukuya was supported by research fellowships from JSPS for young scientists and from Nagoya University program "Leadership Development Program for Space Exploration and Research" for leading graduate schools.

Yeates, A. R., Nandy, D., and Mackay, D. H.   2008, ApJ, 673, 544

Table 1: Leading hemisphere for each sunspot cycle in the past 180 years. The "N" marks indicate that the northern hemisphere led the cycle and the "S" marks indicate that the southern hemisphere did. The "Same" marks represent cases where phase lag hardly existed.

| Cycle No. | Newton, & Milsom 1955 | Waldmeier 1971 ($S$) | Waldmeier 1971 ($D$) | Li 2009 (Group) | Li 2009 (Area) | Murakozy, & Ludmany 2012 |
|---|---|---|---|---|---|---|
| 8  | S   | ... | ... | ... | ...  | ... |
| 9  | N   | ... | ... | ... | ...  | ... |
| 10 | S   | S   | S   | ... | ...  | ... |
| 11 | S   | S   | S   | ... | ...  | ... |
| 12 | N   | N   | N   | N   | N    | N |
| 13 | N   | N   | N   | N   | N    | N |
| 14 | N   | N   | N   | N   | Same | N |
| 15 | N   | N   | N   | N   | N    | N |
| 16 | S   | S   | S   | N   | Same | S |
| 17 | S   | S   | S   | S   | Same | S |
| 18 | S   | S   | S   | S   | S    | S |
| 19 | ... | S   | S   | S   | S    | S |
| 20 | ... | N   | N   | N   | N    | N |
| 21 | ... | ... | ... | N   | N    | N |
| 22 | ... | ... | ... | N   | N    | N |
| 23 | ... | ... | ... | N   | N    | N |

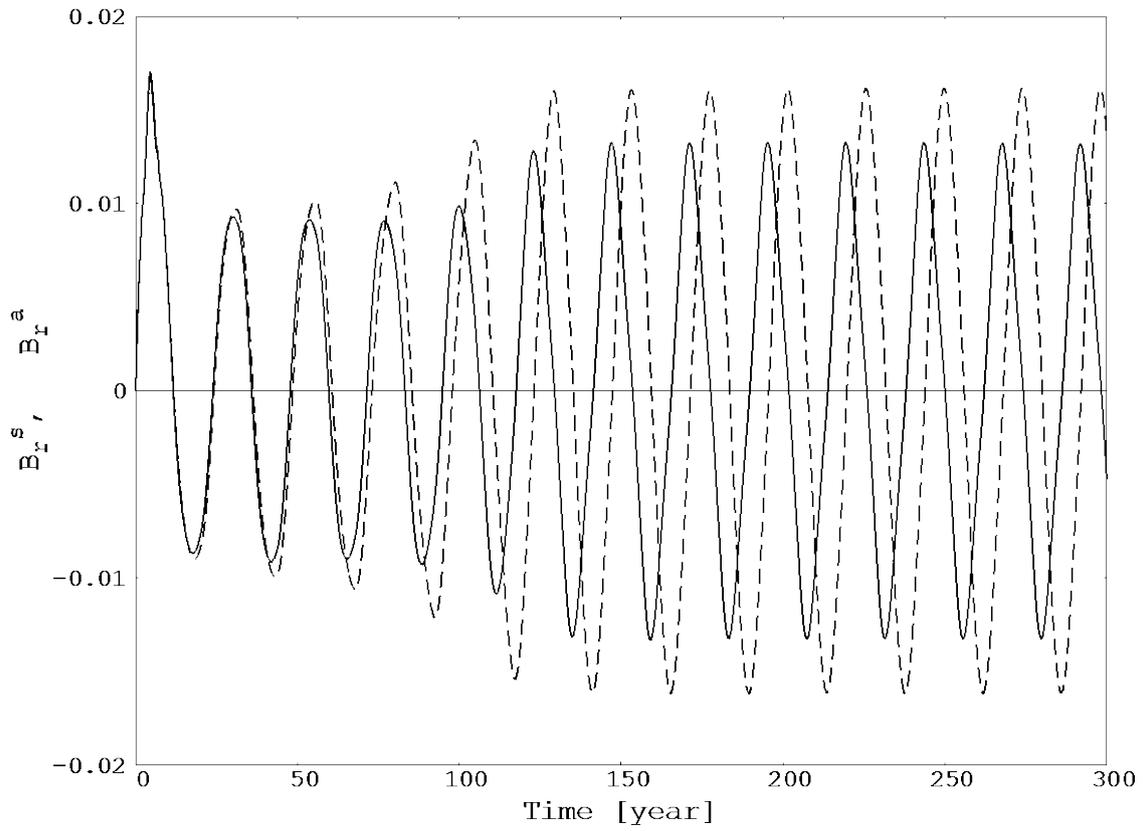

Figure 1: Time evolution of the radial magnetic fields $B_r$ for the symmetric ($B^s$, solid line) and anti-symmetric (B^a, dashed line) components at the north pole. The initial conditions of the amplitude and phase for each component are the same value in this case, where $\eta_{SCZ} = 2.6\times10^{12}$ cm²s⁻¹.

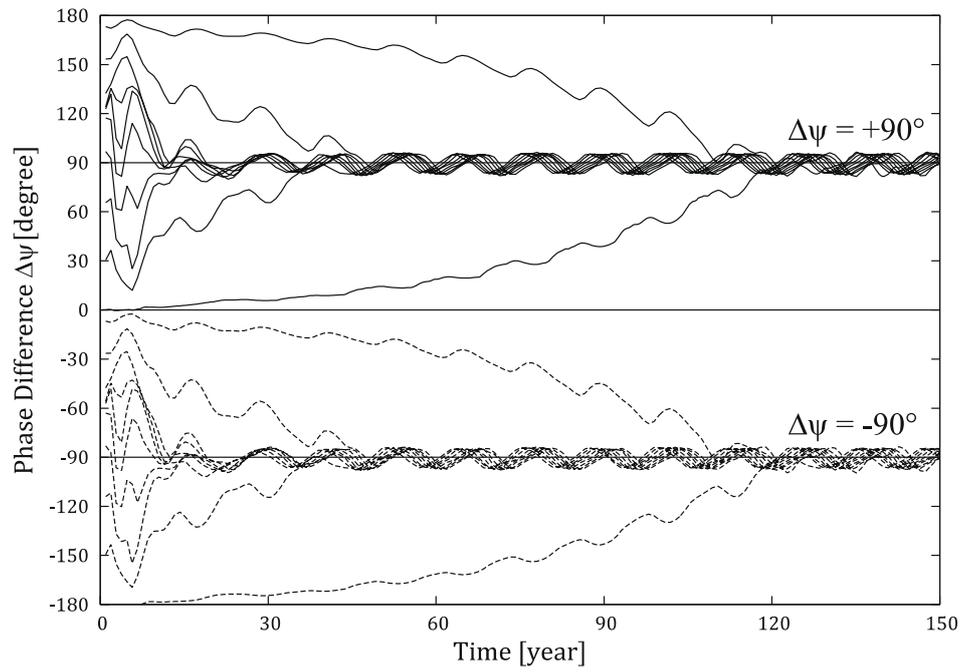

Figure 2: Time evolution of the phase difference $\Delta\psi$ for the 22 different runs of various initial conditions within the range $-180° \leq \Delta\psi \leq 180°$.

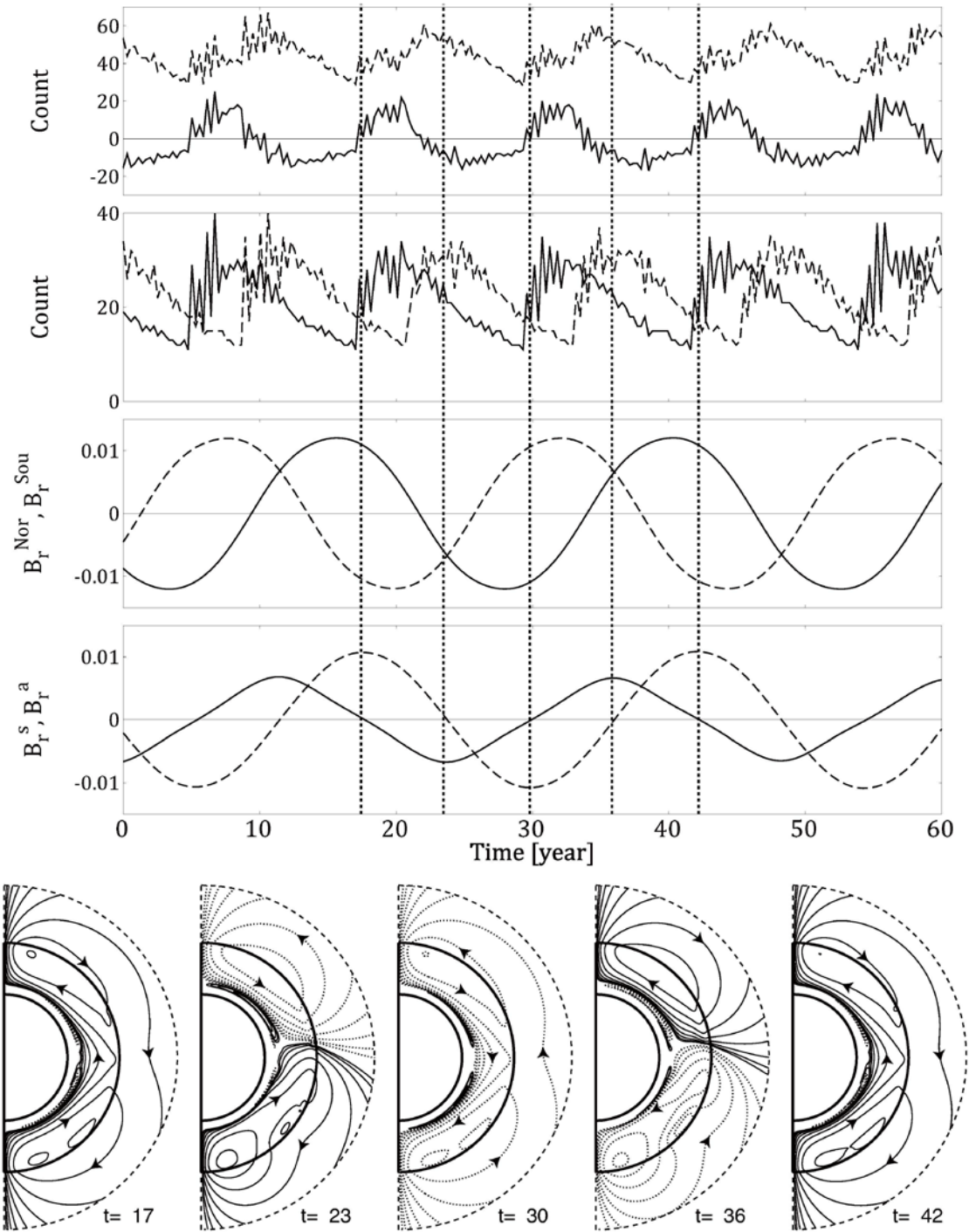

Figure 3: Time evolution of the magnetic field after falling into the attractor for $\Delta\psi = 90°$. Top panel: (dashed line) the total number of grids where the toroidal magnetic field exceeds the critical intensity for the magnetic buoyancy effect: (solid line) the difference in grid number for magnetic buoyancy effects operated between the northern and southern hemispheres. Second panel: the grid number of magnetic buoyancy in the northern (solid

lien) and southern (dashed line) hemispheres. Third panel: the time evolution of the radial magnetic field $B_r$ at the north pole $B_r^{Nor}$ (solid line) and at the south pole $B_r^{Sou}$ (dashed line). Fourth panel: the time evolution of the radial magnetic field $B_r$ of the symmetric (solid line) and anti-symmetric (dashed line) components at the north pole as in Figure 1. Bottom contour maps: the contour of the vector potential $A$ at the times corresponding to the vertical dashed lines in the top three panels. The solid and dashed contours represent magnetic field lines directed by arrows. The portion bounded by thick line corresponds to the simulation box. The contours of potential field in outside of the simulation box are also plotted.

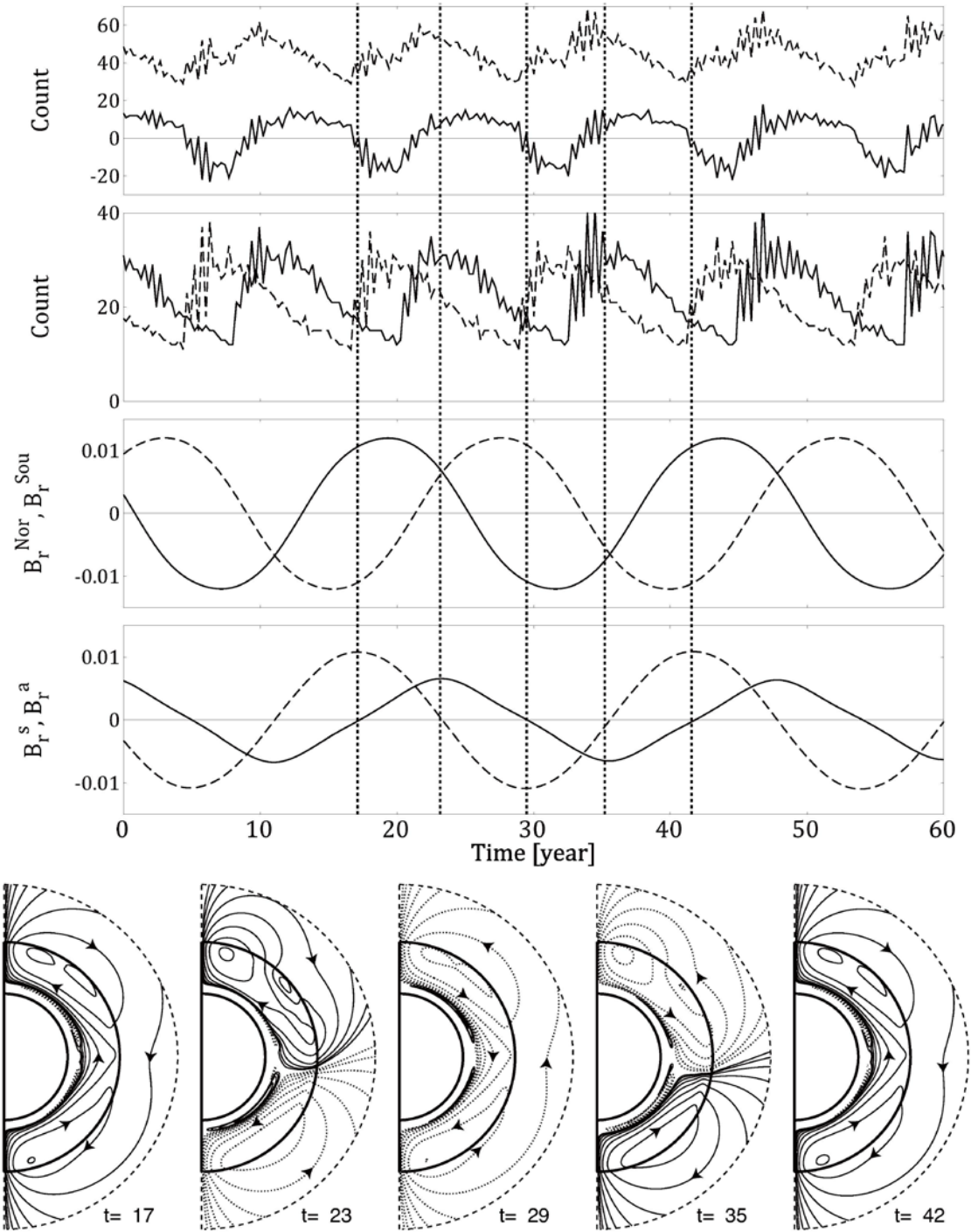

Figure 4: Time evolutions of the magnetic field after falling into the attractor for $\Delta\psi = -90°$. The descriptions for each panel and the contour maps are the same as those in Figure 3.

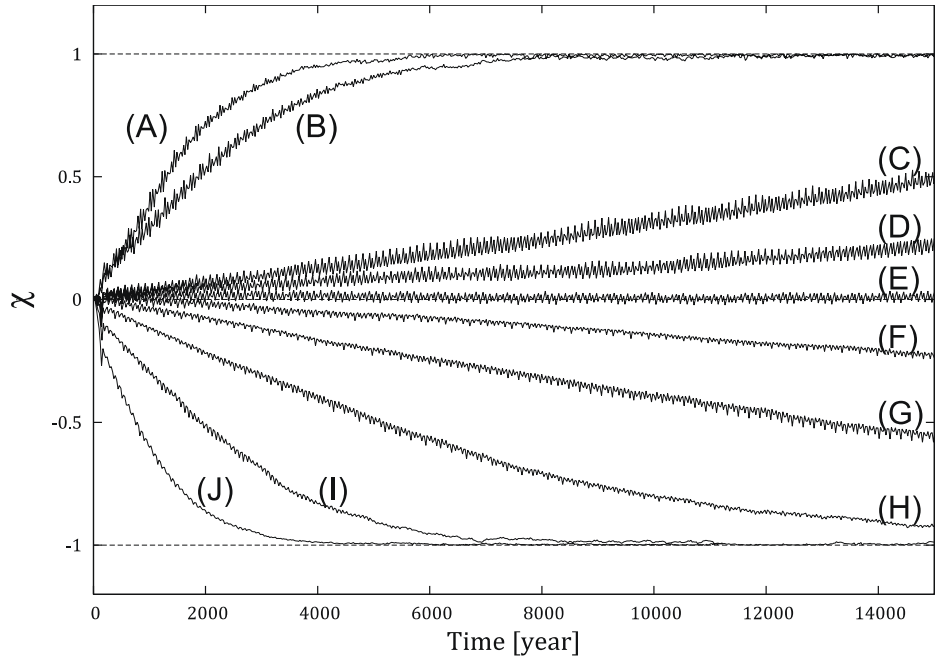

Figure 5: Time evolution of the magnetic ratio $\chi$ at different $Rm$ values. (A) $Rm = 7.65 \times 10^3$, (B) $Rm = 6.80 \times 10^3$, (C) $Rm = 6.12 \times 10^3$, (D) $Rm = 6.09 \times 10^3$, (E) $Rm = 6.06 \times 10^3$, (F) $Rm = 6.03 \times 10^3$, (G) $Rm = 5.97 \times 10^3$, (H) $Rm = 5.83 \times 10^3$, (I) $Rm = 5.56 \times 10^3$, and (J) $Rm = 5.10 \times 10^3$, respectively.

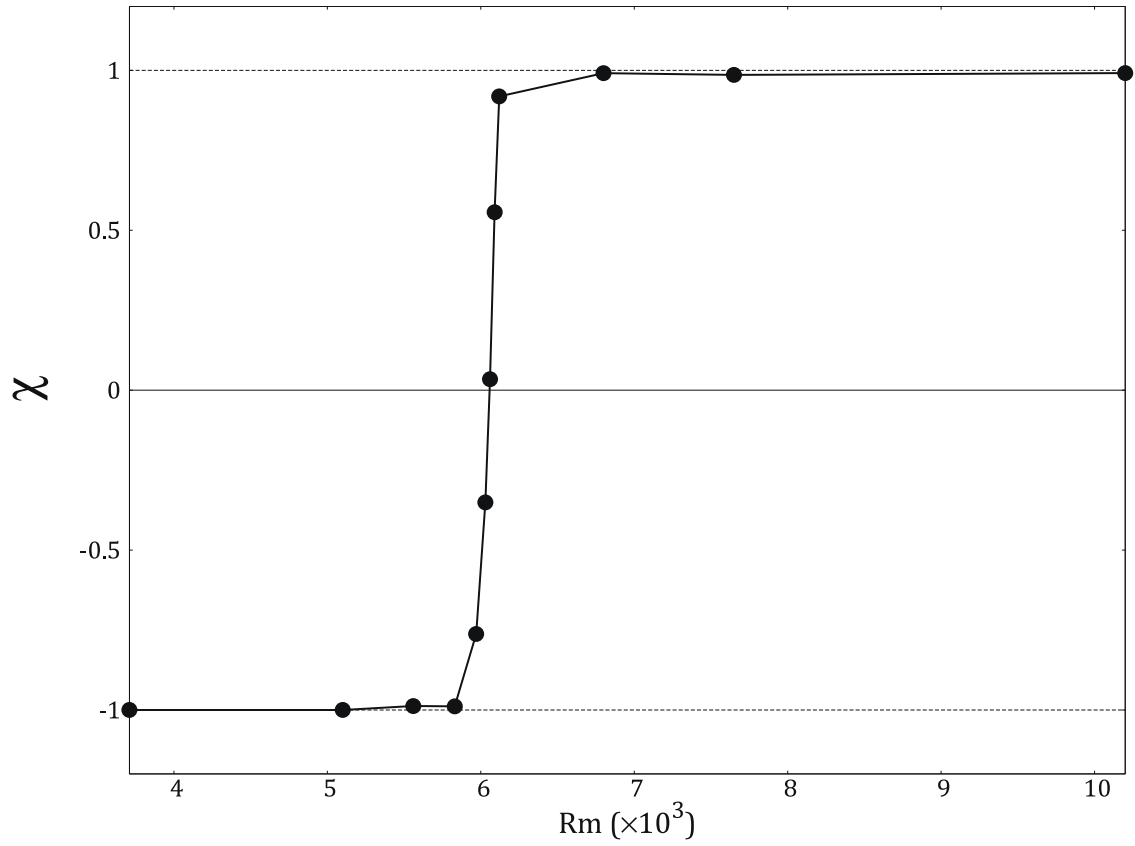

Figure 6: Relation between the magnetic Reynolds number $Rm$ and the ratio $\chi$. The dots represent the data points where we implemented the simulation.

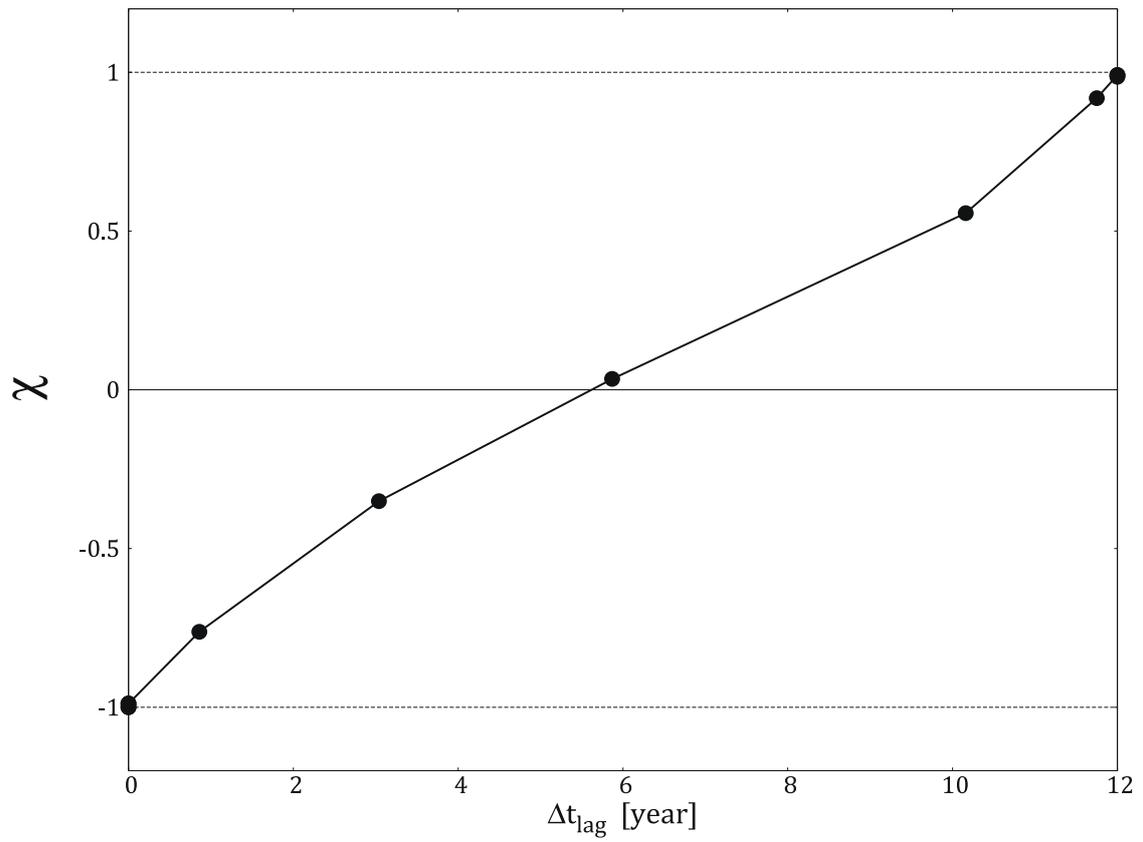

Figure 7: Relationship between the lag time $\Delta t_{lag}$ and the ratio $\chi$. The dots represent the data points where we implemented the simulation.

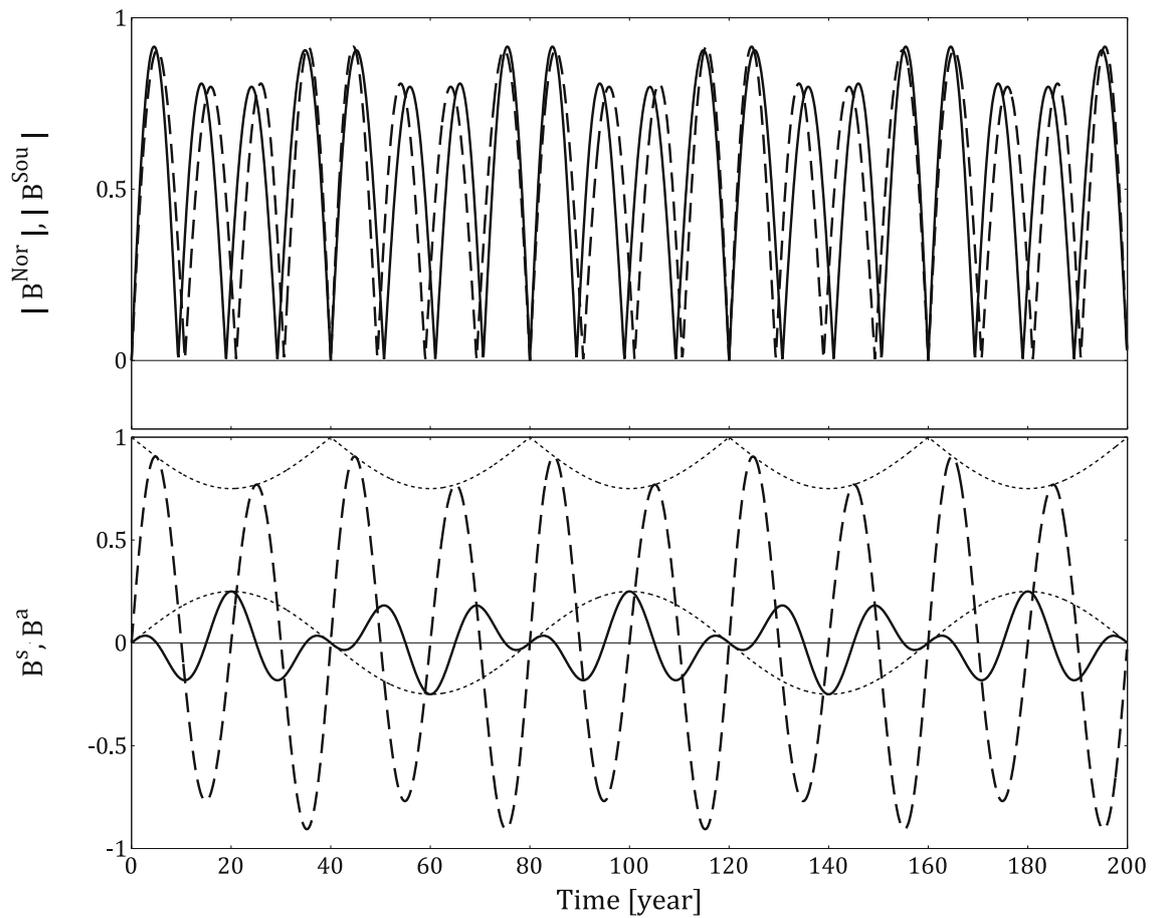

Figure 8: Mathematical experiment for the solar cycle modulation. Upper panel: Variations of the absolute value of the magnetic field on the northern (solid) and southern (dashed) hemispheres. Lower panel: Variations of the symmetric (solid) and anti-symmetric (dashed) magnetic components. The envelope curves for each magnetic component are also plotted (dotted lines).